\documentclass{pasa}
\usepackage{enumerate}
\usepackage{graphicx}
\usepackage{multicol}
\usepackage{fixltx2e}
\usepackage{hyperref}
\usepackage{siunitx}
\usepackage{wrapfig}
\usepackage{lscape}
\usepackage{longtable}
\usepackage{caption}
\usepackage{gensymb}
\usepackage{rotating}
\usepackage{pdflscape}

\title[ADVS]{A digital video system for observing and recording occultations}
\author[Barry et al.]{M.A.(Tony) Barry$^{1}$, Dave Gault$^{2}$, Hristo Pavlov$^{3}$, William Hanna$^{4}$, Alistair McEwan$^{1}$ \and Miroslav D. Filipovi\' c$^{5}$ \\
\affil{$^1$University of Sydney, Electrical and Information Engineering Department, Camperdown, NSW 2006 Australia}
\affil{$^2$Kuriwa Observatory (MPC E28) 22 Booker Rd, Hawkesbury Heights, NSW 2777 Australia}
\affil{$^3$Tangra Observatory (MPC E24) 9 Chad Place, St. Clair, NSW 2759 Australia}
\affil{$^4$190 Gleneagles Trail, Columbia Falls, MT 59912-4390, USA}
\affil{$^5$University of Western Sydney, Locked Bag 1797, Penrith South, NSW 1797 Australia}}

\jid{PASA}
\doi{10.1017/pas.\the\year.xxx}
\jyear{\the\year}

\begin{document}
\begin{abstract}
Stellar occultations by asteroids and outer solar system bodies can offer ground based observers with modest telescopes and camera equipment the opportunity to probe the shape, size, atmosphere and attendant moons or rings of these distant objects. The essential requirements of the camera and recording equipment are: good quantum efficiency and low noise; minimal dead time between images; good horological faithfulness of the image time stamps; robustness of the recording to unexpected failure; and low cost. We describe the Astronomical Digital Video occultation observing and recording System (ADVS) which attempts to fulfil these requirements and compare the system with other reported camera and recorder systems. Five systems have been built, deployed and tested over the past three years, and we report on three representative occultation observations: one being a 9$\pm$1.5 second occultation of the trans-Neptunian object 28978 Ixion (\textit{m}\textsubscript{v} =15.2) at 3 seconds per frame; one being a 1.51$\pm$0.017 second occultation of Deimos, the 12~km diameter satellite of Mars, at 30 frames per second; and one being a 11.04$\pm$0.4 second occultation, recorded at 7.5 frames per second, of the main belt asteroid, 361 Havnia, representing a low magnitude drop ($\Delta$\textit{m}\textsubscript{v} = $\sim$0.4) occultation. 
\end{abstract}
\maketitle%

\section{INTRODUCTION }
 \label{sec:intro}

The observations of asteroid occultations provide the most accurate method to ascertain the size, shape and position of these faint, distant bodies, outside of direct observations by visiting spacecraft. This is particularly true for occultations of trans-Neptunian objects (TNOs) \cite{UltraCamPaper}.

The most frequent asteroidal occultations are of dim stars, because there are more dim stars than bright stars in the night sky \cite{PICOpaper}. Also, the duration of the occultation tends to be on the order of a few seconds, up to a few tens of seconds for the slower-moving TNOs. The magnitude drop during the occultation can be several magnitudes, particularly for TNOs \cite{PHOTpaper}.

Prediction of the best position on Earth to view the asteroid occultation event depends on the combination of the accuracy of the position of the star and the asteroid orbit elements \cite{QuaoarOccPaper}. TNO predictions can have large resultant terrestrial position uncertainties, on the order of thousands of kilometres, and errors in time can be tens of minutes \cite{SolarSystemOccPaper}.

Occultations are distributed widely across the Earth's surface, and sometimes must be observed from remote areas \cite {POETSpaper}. Given these constraints, we consider an occultation observing and recording system should optimise several parameters.

A high quantum efficiency, low read noise, high bit depth camera is required for imaging dim target stars using portable telescopes. The high positional uncertainties of TNO predictions require systems which can be made with minimal necessary cost, to allow for multi-station deployment. 
	
To maximise the usefulness of data collected, the imaging cadence should be as rapid as possible and with minimal inter-image dead time. The system should provide accurate and precise timing, referred to coordinated universal time (UTC) and geolocation information for the images collected.
	
Finally, the system should not have an excessive weight budget, to allow for personal carriage aboard aircraft. Its data storage and retrieval method should be robust to failure and be accessible with common tools for analysis.

\section{PRESENT SYSTEM DESCRIPTION}
\label{sec:presentSystemDescription}

The Astronomical Digital Video occultation observing and recording System (ADVS) is intended to satisfy the requirements presented above.

ADVS is configured to use one of three off-the-shelf Firewire digital video cameras made by Point Grey Research (PGR) (Richmond, BC, Canada), which are described in Section~\ref{sec:testedCameras}. The timing method used to ensure reliable UT-referenced timestamps for each image relies upon specific PGR proprietary features, thus other Firewire cameras may not work with the system.

The camera connects to a laptop or desktop computer, with power and data connection from the camera to the PC using a single Firewire 800 (IEEE1394b) cable. The camera free-runs at an imaging cadence between 30 frames per second (fps) and 8 seconds per frame (spf) with no dead time between images; it sends image acquisition start and stop signals by a separate connector and cable to a custom built GPS-referenced, Hardware Timer and Camera Controller (HTCC) which communicates the time stamps with the PC via USB. 

The timing resolution of HTCC is designed to be within 100~$\mu$s of UTC in the worst case scenario; images are formally timestamped to the millisecond, while our optical timestamp testing system has a 2~ms resolution \cite{SEXTApaper}. The timer also provides geolocation data and GPS satellite fix quality information to the computer via USB.

The computer runs under the open source operating system \textsc{Linux Ubuntu} 10.04 or 12.04, with a bespoke Astronomical Digital Video Recorder program (\textsc{advr}) to display the camera images, apply the timestamps provided by HTCC to the images, and to write the images, data, and metadata (e.g. camera parameters, geolocation and satellite information) to disk as a single file structure. 

Computing requirements are modest - we have used ADVS with a \url{~}\$200 PC running a Via Nano Processor (2~core; 1.6~GHz processor, 4GB~RAM @ 533~MHz FSB) and a 60~GB solid-state drive without issues. The main requirement is that IEEE1394b must be supported by some means, whether on the motherboard, via a PCIe card, an Express/34 slot or Thunderbolt adapter. 

\section{ADVS SCHEMA}
\label{sec:ADVSoperation}

\subsection{CAMERA}
PGR provide several proprietary additions to regular camera behaviour which allow \textsc{advr} to reliably timestamp images. Each image can have metadata embedded into the actual image itself, in the first few (Row 0, Pixels 1 to 40) pixels.  Each metadata pixel carries a value, which represents frame number, a time stamp based on the Firewire bus clock, as well as other image data parameters like gain, shutter duration, etc.

The embedded image metadata provides the reference to allow HTCC timing data to be matched with the correct image.

\subsection{HTCC}
HTCC has a GPS unit which outputs one-pulse-per-second (1PPS) signal referred to UTC to within 1~$\mu$s, and associated NMEA statements which describe the signal. This provides the discipline for the timing function of HTCC and provides observer site geolocation. 

PGR cameras offer a hardware output signal which goes low on camera imaging commencement, and goes high at image end.  HTCC can therefore precisely measure the actual image start and end time, and forward the information to \textsc{advr} via USB. The cameras also have an input trigger which allows for hardware initiated image acquisition. HTCC is able to unambiguously trigger a singular first image of a known image duration, which provides the zero reference for future image acquisitions. 

\subsection{ADVR}
\textsc{advr} has a number of features to assist the operator of the recorder. The display normally runs in a window displayed over part of the PC screen. It can be expanded to cover the whole screen and eliminate other display element distractions (e.g menu bar, other windows, etc). 

The time (UTC), geolocation and status indicators normally are displayed on the window. They can be hidden so only the star field is displayed. See Figure 1 for the \textsc{advr} window and annunciators. 

Image time is sent from HTCC as start and end timings, and \textsc{advr} computes the central time and duration and displays this on-screen.

The computer screen displays linearly mapped luminosity images (camera to screen). However, the mapping can be changed to enhance low-brightness objects near the noise floor. The screen can also be inverted (black stars on white). We found that these features assist the operator in detecting faint stars which would otherwise not appear on-screen because of the limitation of the bit depth of LCD screens. Frame rate, camera gain, offset and gamma are changed by keyboard entry and the settings are displayed on screen while being changed.

\textsc{advr} records images and metadata (timestamp, image duration, camera settings, geolocation information etc) to disk in a lossless \textsc{.adv} file format, which has been crafted to ensure that if sudden failure (e.g power outage) occurs, the images and metadata can be recovered up to the point of failure. This is done by writing the file header and data sections definition to disk at file creation time, providing an index into the rest of the file's image and metadata data blocks. Each subsequent data block is then written when required, with a predefined number as the first four bytes. If an error occurs and the file write is terminated abruptly without proper closure, the recovery utility examines the header and definitions file to determine the general size of each data block and then continues through the rest of the file on disk looking for the predefined number to confirm the next block's start point in the file. 

The \textsc{.adv} file can be exported to \textsc{fits} images for external analysis. The file can also be examined without conversion in \textsc{Linux}, \textsc{Windows} or \textsc{Mac OS} by the \textsc{Tangra v3} application which has been developed by one of the authors (H.P.) to allow light curve reduction from the \textsc{.adv} file \cite{TangraWebSite}. The \textsc{adv} file format is available online \cite{ADVGitHub}.

%____________Fig. 1
\begin{figure}[h!]
\begin{center}
\includegraphics[width=\columnwidth]{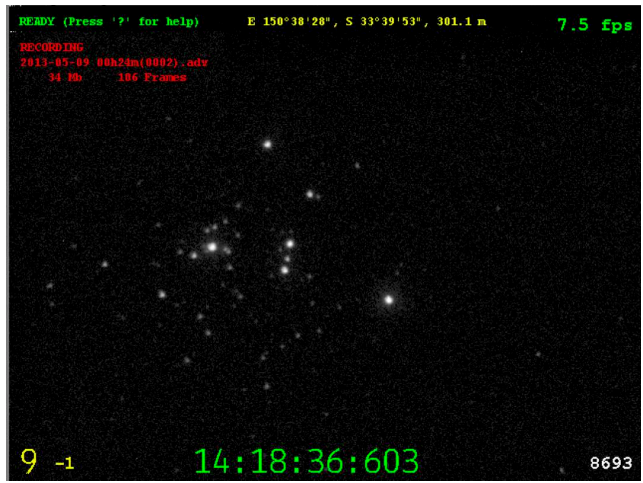}
\caption{\textsc{advr} screen elements. Clockwise from top left - Ready/Recording annunciator and file size; geolocation data; frame rate indicator; frame counter reading 8693 frames from start and dropped frames counter (not shown here because there were none); Universal time, GPS-UTC firmware adjustment (--1~sec); number of GPS satellites in fix.}
 \label{Fig1}
\end{center}
\end{figure}
%__________________

\section{TESTED CAMERAS}
\label{sec:testedCameras}

The PGR cameras tested were all monochrome \mbox{C-Mount} video cameras with square pixels. The imaging mode was progressive (continuous frame), rather than interlaced (two half-height, interdigitated fields per frame), such as would be found on analog video cameras. Progressive mode greatly simplifies timing issues, as the frame is all exposed at once, rather than having the odd CCD lines exposed in one field first, then the even lines exposed in the next field.

The cameras operated in free-run cadences which ranged from 30~fps down to 8~spf in 12 steps. The Flea cameras were able to be operated at 60~fps but this was not used for occultations described here.

A test of the cameras with the star cluster M\,7 as the target was recorded on 7$^{th}$ February 2012. The three target stars within M\,7 are described in Table~\ref{starTestTable}. 

The settings used for all cameras was 1.28 seconds duration, gain at max for each camera and gamma at 1 (off). Stars were near the zenith at the time and the expected magnitude difference over the time of the test due to air mass change or atmospheric extinction change was assessed using the ICQ procedure to be less than 1 milli-magnitude \cite{ICQpaper}. The telescope was a 350~mm Meade LX200 ACF fitted with a f/3.3 focal reducer.
 
\begin{figure*}[h]
\begin{center}
\includegraphics[width=\textwidth]{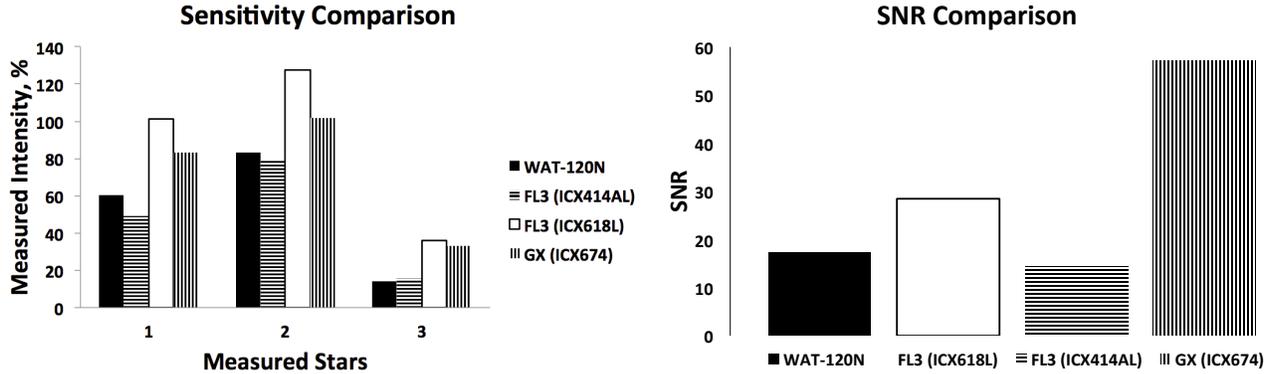}
\caption{Sensitivity and SNR comparison for target stars in M\,7. For details see Section~\ref{sec:testedCameras}}
\label{sensitivityFig}
\end{center}
\end{figure*}

Sensitivities of the ADVS cameras, and a Watec 120N analog interlaced videocamera are shown in Figure~\ref{sensitivityFig}. This is a comparison of measured star intensities for each of the three target stars. An intensity of 100\% is classed as a ``full well'' pixel measurement of 255 for the 8-bit Watec 120N, 4095 for the 12-bit Flea cameras and 16383 for the 14-bit Grasshopper Express camera. If a stellar disc spans more than one pixel, the total intensity can be greater than 100\%.

A comparison of signal to noise ratio (SNR) is also shown in Figure~\ref{sensitivityFig}. The noise in this figure is the 1-$\sigma$ error in the three star intensity measurements (rather than the background noise derived from the images). Camera details are summarised in Table~\ref{cameraTable}.

The Watec camera has been used previously in occultation recordings \cite{CharonOcc2008Paper} \cite{ProAmPaper} and is provided so comparison with a known instrument can be made. 

The Grasshopper Express (GEx) had the lowest read noise (9.6e-), the largest chip area (7.3$\times$5.5~mm), the highest bits-per-pixel (14 bits) and the highest cost (\$1990). All the occultations observed in this article used the GEx camera.

\begin{table*}[h!]
\caption{Star targets in M\,7}
\label{starTestTable}
\begin{center}
\begin{tabular*}{\columnwidth}{@{}l l l l l r@{}}
\hline
No.& UCAC4& Bmag& Vmag& Notes\\
\hline
1& 4U 347-016724& 12.929& 12.534& Bright star\\
\noalign{\vspace {.2cm}}
2& 4U 347-016625& 12.934& 11.771& Red star\\
\noalign{\vspace {.2cm}}
3& 4U 347-016673& 14.346& 13.861& Faint star\\
\hline 
\end{tabular*}
\end{center}
\end{table*}

\begin{table*}[h]
\caption{Camera details}
\label{cameraTable}
\begin{center}
\begin{tabular*}{41.5pc}{@{}l l l l l r@{}}
\hline
& Small Flea& Big Flea& Grasshopper Express& Watec\\
\hline
Camera model& FL3-03S1M-C & FL3-03S3M-C& GX-FW-28S5M-C& 120N\\
\noalign{\vspace {.2cm}}
CCD, format& 
ICX618, 640$\times$480 & 
ICX414, 640$\times$480 & 
ICX674, 1600$\times$1200 2$\times$2 bin&
Sony*, 768$\times$495 \\
Native pixel size&
5.6~$\mu$m$^2$&
9.9~$\mu$m$^2$&
4.54~$\mu$m$^2$ \# &
8.4~$\times$~9.8$\mu$m \\
\noalign{\vspace {.15cm}}
Frame rate& 8 spf to 60 fps& 8 spf to 60 fps& 8 spf to 30 fps& 8.3 spf to 30 fps\textdagger \\
\noalign{\vspace {.15cm}}
Dead time& $<$2 msec& $<$2 msec& $<$2 msec& $<$ 2 msec\\
\noalign{\vspace {.15cm}}
Read noise& 39e-& 63e-& 9.6e- & not stated\\
\noalign{\vspace {.15cm}}
Dark current $@$ 20\degree C& 288e-/sec& 256e-/sec& 1.8e-/sec & not stated\\
\noalign{\vspace {.15cm}}
Quantum efficiency& 67\%& 44\%& 67\% & not stated\\
\noalign{\vspace {.15cm}}
Bits, ADU count& 12 bit, 4096& 12 bit, 4096& 14 bit, 16384 & 8 bit, 256\\
\noalign{\vspace {.15cm}}
Well depth& 23035 e-& 18292 e-& 15800 e-& not stated\\
\noalign{\vspace {.15cm}}
Cost& US\$ 500&  US\$ 900& US\$ 1990 & US\$ 600 \\
\hline 
\end{tabular*}
\end{center}
\tabnote{* Analog sensor, type not specified by Watec, horizontal resolution dependant on frame grabber.}
\tabnote{\# native CCD is 1932$\times$1452 but is used in a reduced mode to deliver a 800x600 field with 9.08~$\mu$m$^2$ pixels. }
\tabnote{\textdagger~60 fields per second are possible with interlaced CCDs, but only half of the CCD is exposed in any field.}
\end{table*}

\section{TIMING VERIFICATION}
\label{sec:timingVerification}
Imaging time stamp verification required us to build a device (``SEXTA'') which provides an optical indication of the passage of time. SEXTA has been described previously \cite{SEXTApaper}.

The SEXTA display was placed in the field of view of the ADVS system, and each frame of the camera output was compared, optical timestamp from SEXTA against the internal timestamp in the image header from ADVS, for 10\,000 sequential frames at 30~fps. It was found that image optical timestamps did not differ from header timestamps to the limits of SEXTA resolution (2~msec precision, fiducial to UT to within 0.2~msec). Dead time between exposures for all frame rates was found to be less than SEXTA resolution ($<$ 2~msec).

\section{OBSERVED OCCULTATIONS}
\label{sec:observedOccs}
ADVS was used to observe an occultation of the TNO 28978 Ixion from Alice Springs, Australia, on the 24$^{th}$ June 2014 at 14:54~UT, following a prediction by F. Braga-Ribas and the RIO team \cite{PredictionPaper}. The target star was a close double of combined magnitude 15.2. The telescope used was a Meade 200~mm LX90 with a f/3.3 focal reducer, giving a plate scale of 0.96 arcsec/pixel for the camera.

\begin{figure*}[h!]
\begin{center}
\includegraphics[width=17.5cm]{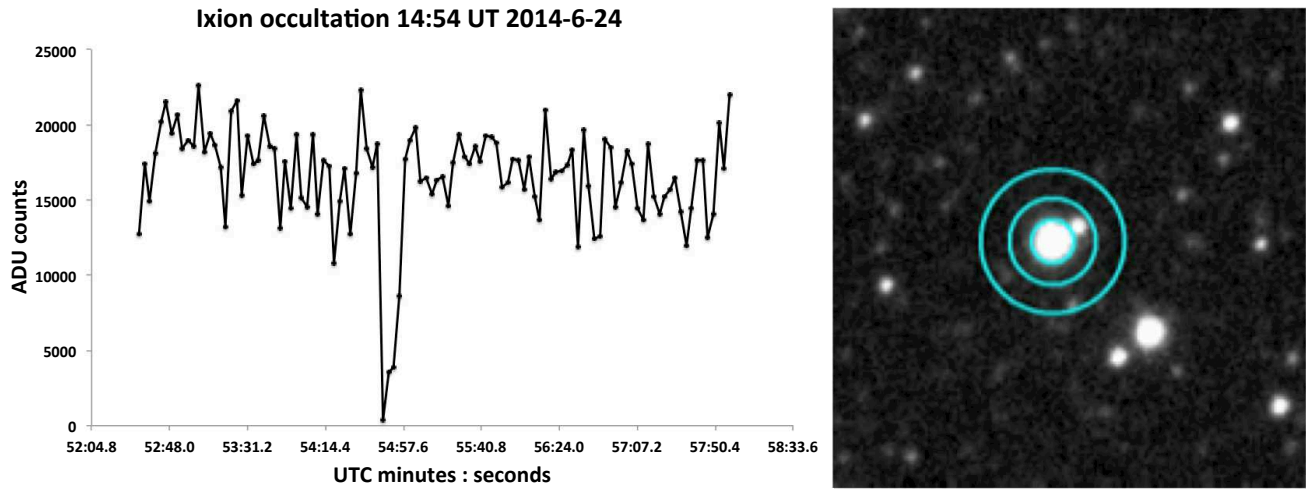}
\caption{Light curve of Ixion occultation, and target star showing close visual double. Image courtesy of Julio Camargo. For details see section~\ref{sec:observedOccs}}
\label{IxionOccFig}
\end{center}
\end{figure*}

The light curve (see Figure~\ref{IxionOccFig}) was reduced with an aperture of 3.3 pixels for the star/TNO. S/N was measured at 6.05:1 with a cadence of 3~spf. The occultation duration was 9.0$\pm$1.5~seconds. The non-zero flux for two points during the occultation was attributed to one star of the close double. 

ADVS was used to observe an occultation of Deimos, the 12~km diameter satellite of Mars, from Alice Springs, Australia on the 17$^{th}$ June 2014 at 11:47~UT following a prediction by J. Broughton (personal communication to W.H., 2014). The telescope and camera were the same as for the Ixion occultation. The target star was HIP 62565 of magnitude 8.9, and Deimos was just 20~arcsec from its mag --0.7 primary (Mars). 

\begin{figure*}[h!]
\begin{center}
\includegraphics[width=17.5cm]{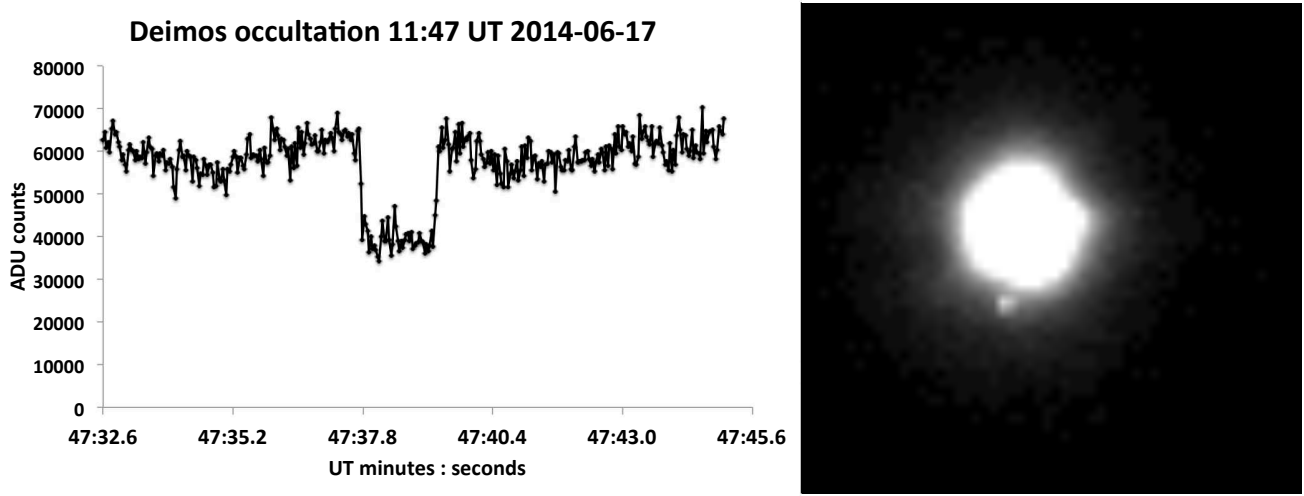}
\caption{Light curve of Deimos occultation. The coalesced HIP62565 + Deimos occupy just 4 pixels in the halo around Mars. For details see section~\ref{sec:observedOccs}.}
\label{DeimosOccFig}
\end{center}
\end{figure*}

The light curve (see Figure~\ref{DeimosOccFig}) was reduced with an aperture of 2.4 pixels for the star/satellite. S/N was measured at 3.23:1 at the time of the disappearance, with a cadence of 30~fps. The occultation duration was 1.510$\pm$0.017~seconds.

\begin{figure*}[h!]
\begin{center}
\includegraphics[width=17.5cm]{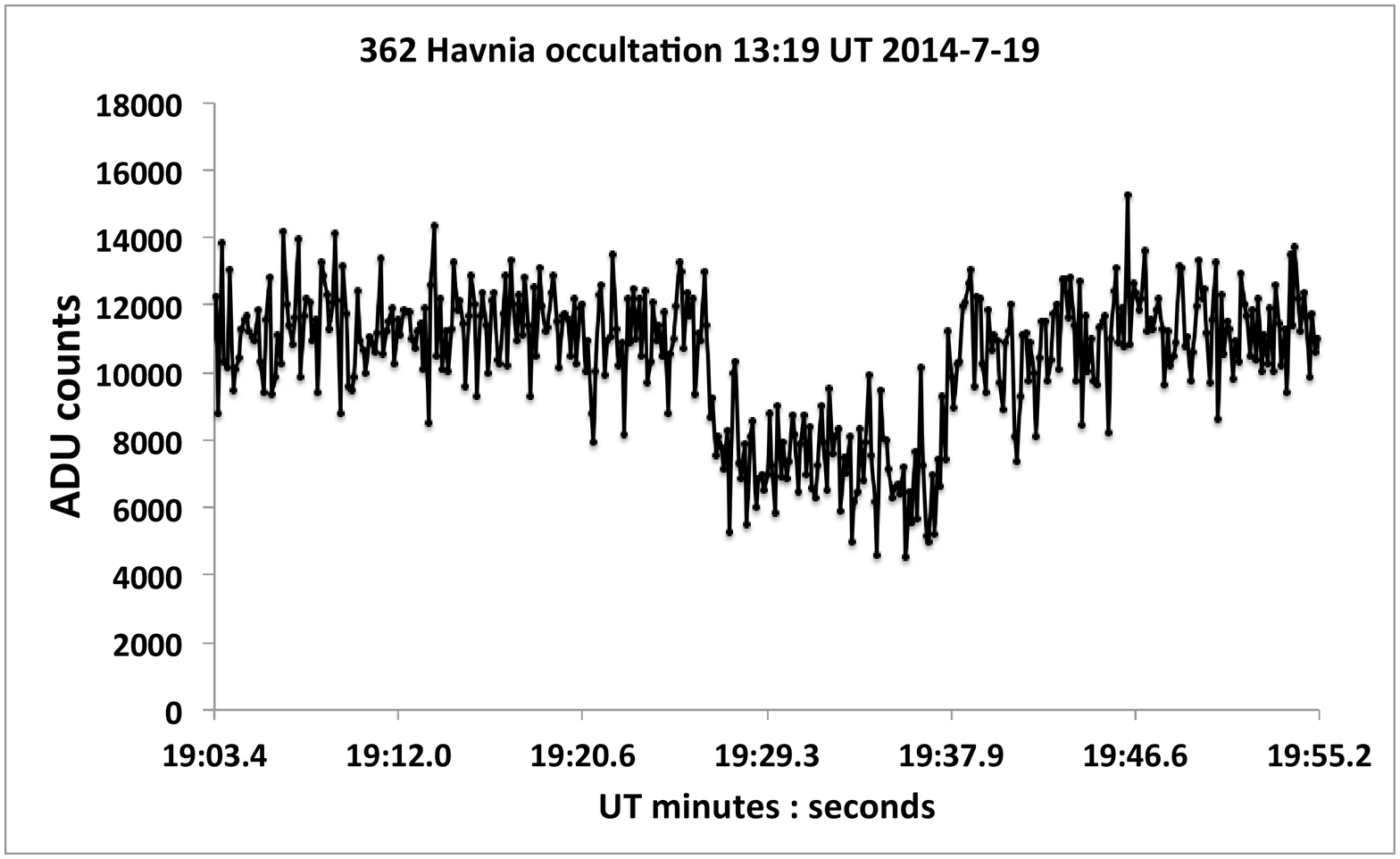}
\caption{Light curve of occultation of 362 Havnia, $\Delta$\textit{m}\textsubscript{v} = \url{~}0.4. For details see section~\ref{sec:observedOccs}.}
\label{HavniaOccFig}
\end{center}
\end{figure*}

ADVS was used to observe a low-magnitude drop occultation of 4UC274-132026 by 362 Havnia, a main belt asteroid, from the Blue Mountains, NSW on the 19$^{th}$ July 2014 at 13:19~UT. The telescope was a 300~mm Meade LX200 with a 0.5$\times$ focal reducer, giving a scale of 1.04~arcsec/pixel for the camera. The target star was \textit{m}\textsubscript{v} = 13.6, with a drop of 0.4 magnitudes. The light curve (see Figure~\ref{HavniaOccFig}) was reduced with an aperture of 3.5~pixels, with S/N measured at 9.2:1 at 7.5~fps. The occultation duration was defined as 11.04$\pm$0.40~seconds, with the larger uncertainty due to the shallower slope of the disappearance and reappearance. This is likely due to atmospheric seeing coupled with a low-mag drop, rather than a large stellar diameter or tangential approach. The chord was near central given Havnia's 98km expected diameter and a $\sim$8.1km/sec shadow velocity. The occulted star's angular diameter was calculated to be 30 micro-arcsec using B and V magnitudes of 14.7 and 13.6 respectively \cite{StellAngDiamPaper} versus Havnia subtending an angle of 68 milli-arcsec.

%_____________Fig. 2__________________________________
%____________________________________________________
%_____________Fig. 3__________________________________
%____________________________________________________

%_____________Fig. 4__________________________________
%____________________________________________________

\section{COMPARISON OF ADVS WITH OTHER RECENT SYSTEMS}
\label{sec:comparisonADVS}

The Portable Occultation, Eclipse, and Transit System (POETS) was developed by Williams College and the Massachusetts Institute of Technology in 2006 \cite{POETSpaper}. The camera is an iXon DU-897 cooled EMCCD frame-transfer device from Andor Instruments (Belfast, UK). The system comprises the camera, a desktop computer rather than a laptop due to the interface card and system resources required, and a commercially available GPS-based controller, the TM-4 from Spectrum Instruments (San Dimas, USA).

The Portable High-Speed Occultation Telescope system (PHOT) was developed by Southwest Research Institute and Wellesley College in 2005 (Regester et al, 2011). The camera was initially a MicroMax 512BFT and later a PhotonMax 512B, both cooled EMCCD-capable devices of 512$\times$512$\times$13~$\mu$m pixels, from Roper Scientific (New York, USA). The system comprises the camera, an interface box in the case of the MicroMax camera, a Dell D620 laptop with dock to house the interface card for the camera, and an in-house-built GPS-based AstroTimer. 

The Portable Instrument for Capturing Occultations (PICO) was developed by Williams College and the Massachusetts Institute for Technology in 2010, as a lower cost alternative to the POETS system \cite{PICOpaper}. The camera is a Finger Lakes Instrumentation (New York, USA) ML261E-25 cooled (but not EM) unit based on the Kodak KAF-0261E CCD. The camera has a mechanical shutter and a cadence of 1~second exposures every 1.76~seconds. The GPS controller is the Spectrum Instruments TM-4, and the PC is a Lenovo IdeaPad S12 laptop which uses M\textsc{axim}DL commercially available software to interface with the camera and timer via USB2.

The system details for POETS, PHOT and PICO are compared with ADVS and summarised in Table~\ref{systemDescriptionTable}.

\begin{table*}
\caption{System description and comparison}
\label{systemDescriptionTable}
\begin{center}
\begin{tabular*}{41.5pc}{@{}l l l l l r@{}}
\hline \hline
& POETS& PHOT& PICO& ADVS\\
\hline
Camera&  Andor & Andor& Finger Lakes Instr.& Point Grey Research\\
& iXon DU-879& PhotonMax 512B& ML261E-25& Grasshopper Express\\
& & & & GX-FW-28S5M-C\\
\noalign{\vspace {.2cm}}
CCD type& Electron & Electron& Kodak KAF-0261E& Sony ICX674\\
& multiplying& multiplying& & \\
\noalign{\vspace {.2cm}}
Cooling& active& active& active& passive\\
\noalign{\vspace {.2cm}}
Image format & 512$\times$512& 512$\times$512& 512$\times$512& 1600$\times$1200\\
pixel size& 16~$\mu$m$^2$& 13~$\mu$m$^2$&  20~$\mu$m$^2$ 2$\times$2 bin& 4.54~$\mu$m$^2$ 2$\times$2 bin\\
\noalign{\vspace {.2cm}}
Best bit depth& 16 bit&  16 bit& 16 bit& 14 bit\\
\noalign{\vspace {.2cm}}
Frame rates& 1 -- 30 fps& 1 -- 20 fps& 0.568 fps& 8spf -- 30 fps\\
\noalign{\vspace {.2cm}}
Max rate format& 180$\times$120 $@$ 30~fps& 64$\times$64 $@$ 20~fps& 512$\times$512 $@$ 0.568~fps& 800$\times$600 $@$ 30~fps\\
\noalign{\vspace {.2cm}}
Interframe time& a few msec\textdagger& a few msec\textdagger& 0.76 sec& $<$2 msec*\\
\noalign{\vspace {.2cm}}
Read noise& 0 (EMCCD)& 0 (EMCCD)& 13.6 e-& 9.6 e-\\
\noalign{\vspace {.2cm}}
Dark current& $\sim$0 e-/sec $@$ -70\degree C& 0.008 e-/sec $@$ -80\degree C& 1e-/sec $@$ -30\degree C& 1.8e-/sec $@$ 20\degree C\\
\noalign{\vspace {.2cm}}
Weight, size& 2$\times$14~kg airline cases& 2 cases 22~kg total& 11~kg airline case& 5~kg carry-on baggage\\
\noalign{\vspace {.2cm}}
Power budget& 
mains operation& 
900~W cooldown& 
mains operation& 
6.75~W cam + timer\\
& 
only& 
300~W maintenance& 
& 
85~W laptop\\
\noalign{\vspace {.2cm}}
Cost& US\$ 38\,000& $\sim$ US\$ 30\,000& US\$ 5\,000& US\$ 4\,500\\
& whole system& camera only& whole system& whole system\\
\hline \hline
\end{tabular*}
\end{center}
\tabnote{\textdagger Interframe (dead) time as stated by article authors.}
\tabnote{*Limits of optical time stamp testing device. Actual values are likely to be lower.}
\end{table*}%

\section{DISCUSSION}
\label{sec:discussion}

Both the POETS and the PHOT systems use EMCCD cameras with near-zero read noise, quantum efficiencies around 90\%, imaging cadences to 20~Hz and a price tag above \$30\,000. While this performance is highly desirable, the cost means that such systems cannot be deployed on a large scale. The small region of interest on the chip (64$\times$64 or 180$\times$120) and resultant tiny field of view is one of the penalties of the EMCCD cameras when used at high frame rates. The PICO system costs around US\$5\,000 and so can be acquired more easily than EMCCD devices, but the low frame cadence of 1.76~sec and inter-frame dead time of 0.76~sec per image constrains the resolution of the asteroidal data collected. Lockhart et al (2010) notes that this cadence would not be sufficient to identify details in the atmospheric region of Pluto, but it would provide evidence of atmosphere.
	
ADVS, when used with the GEx camera provides better quantum efficiency (68\%  vs. 59\%) and lower read noise (9.1e-  vs. 13.2e-) when compared to PICO, with much greater choice in the cadence rate (from 30~fps to 8~spf, in 12 steps), and no measurable inter-frame dead time. The FLI camera has a larger sensor area (105~mm$^2$  vs. 40~mm$^2$) but fewer pixels (262k  vs. 480k) leading to a less detailed Point Spread Function for a given star. The cost of the GEx is \$1\,990, slightly over half of the FLI camera used by PICO. 

Dark current for the POETS, PHOT and PICO systems is vanishingly low due to the deep cooling these cameras offer and the short exposures required for occultation cadences. In the case of PHOT, the dark current is quoted at 0.008e- per second. 

The GEx dark current is quoted at 1.8e- per second at room temperature \cite{GExSpecs}, and this reflects the uncooled nature of the camera. This is still less than 20\% of the read noise for a 1~second exposure, and demonstrates the fact that for occultation work, read noise is the major determining camera handicap (Regester et al, 2011). Dark current halves for every 7 degrees Celsius drop in temperature, so case-cooling the GEx to near zero degrees Celsius would drop the dark current to about 0.6e- per second. This is an area of future investigation.

The GPS controllers used by POETS, PICO and PHOT are all of the ``push'' type - they output a pulse or pulse train which requires the camera to complete its imaging cycle (expose, process, download) within that pulse period \cite{MORISpaper}. This has limitations as the pulse cadence must be at least as long (or longer) than required by the camera to complete all operations or there will be an imaging cadence break. This might be an immediate dropped frame or the frame start time might become progressively more and more delayed until a dropped frame occurs. Exactly which scenario occurs depends on the camera characteristics, however, neither are desirable. If a trigger is dropped and does not cause the camera to generate an image, it becomes difficult to determine where the gap in the cadence occurred. The PC system clock time will often be included with image creation and may be able to help if the cadence is more leisurely than 1~fps. For faster cadences, the hole created by a singular dropped image could be difficult to identify without additional image tagging.

In contrast, HTCC measures the free-run cadence of the camera and provides the start and end time of each exposure to the recorder. The camera keeps its own imaging cycle operating at optimal efficiency and minimum dead time and the timer measures the frame start and stop at the time they actually occur. 

The power requirements of the four camera systems discussed (POETS, PHOT, PICO and ADVS) varied widely due primarily to the necessity of Peltier cooling for the cooled cameras. The PHOT system paper gave a breakdown of the power budget for the system, which was around 900~W on cool-down, and 300~W on maintain. The POETS and the PICO systems paper did not provide power requirements, except to say that the system had universal supplies to allow for non-USA operation from mains power.

ADVS requires 5~W for the GEx camera which is supplied by the FireWire bus of the laptop and 1.75~W for HTCC, which is supplied by the USB bus of the laptop. A typical Macintosh laptop has an 85~W power adapter and therefore this is the power requirement for the ADVS system. As a bonus, the laptop battery can drive ADVS for more than an hour with no AC power applied.

The three occultations reported here, observed with ADVS, were all accomplished with the aid of telescopes of apertures 300~mm or less, which is a ``telescope of opportunity'' rather than a fixed installation \cite{PHOTpaper}. These are portable instruments and hence of small apertures. In the case of the Ixion occultation, the telescope aperture was 200~mm, the star's visual magnitude was 15.2, and the cadence was 3 seconds per frame with less than 2~msec interframe time, for a SNR of 6.05. The main requirement was for a relatively long exposure (3~sec) with low noise and minimal dead time to make up for the limited aperture of the instrument and the faint target. 

In the case of the Deimos occultation, the glare from Mars at 20~arcsec separation and 30~fps represents a different requirement (high frame rate) in the camera and recording system. For the Havnia occultation, the main requirement was for low noise and high bit depth at a moderate rate (7.5~fps).

\section{CONCLUSION}
ADVS is a portable and low cost occultation recording system which meets the requirements of TNO occultation recording. At present, it is the only digital occultation recording system known to the authors that measures exposure time and duration directly from the camera, rather than via assumption of action by a push signal from a controller. Its timebase is synchronised to within 2~msec of UT by GPS referencing. Dead time between images is less than 2~msec. Frame rates from 8~spf to 30~fps are supported.

ADVS has been used to record a TNO occultation, several main belt asteroidal occultations and an occultation of Deimos, the smaller moon of Mars. Five development systems are currently in use.

\section{ACKNOWLEDGEMENTS}
The field image in Figure~\ref{IxionOccFig} is courtesy of Julio Camargo $<$camargo$@$on.br$>$, via a personal communication from Jose-Luis Ortiz $<$ortiz$@$iaa.es$>$. The image was obtained with the ESO/MPG 2.2-m telescope during run 082.A-9211(A), within the ESO - Observatorio Nacional/MCTI agreement. T.B. was supported by a Joint Research Engagement (JRE) grant from the University of Sydney (2012 -- 2014). D.G. received a donation of one Grasshopper Express camera from Point Grey Research for testing.

%\section*{CONFLICTS OF INTEREST}
%None.


\begin{thebibliography}{}

\bibitem[\protect\citename{Assafin et al., 2012}]{PredictionPaper}
Assafin M., Camargo J.I.B., Viera Martins R. et al., 2012 A\&A, 541:A142 

\bibitem[\protect\citename{Barry et al., 2015}]{SEXTApaper}
Barry M.A., Gault D., Bolt G. et al., 2015 PASA, 32, e014

\bibitem[\protect\citename{Braga-Ribas et al., 2013}]{QuaoarOccPaper}
Braga-Ribas F., Sicardy B., Ortiz J.L. et al., 2013 AJ, 773:26

\bibitem[\protect\citename{Dhillon et al., 2007}]{UltraCamPaper}
Dhillon V.S., Marsh T.R., Stevenson M.J et al 2007, MNRAS 378 (3): 825-840

\bibitem[\protect\citename{Green, 1992}]{ICQpaper}
Green D.W.E., 1992, International Comet Quarterly, 14:55-59

\bibitem[\protect\citename{Gulbis et al., 2011}]{MORISpaper}
Gulbis A.A.S., Bus S.J., Elliot J.L. et al., 2011, PASP, 123, 902

\bibitem[\protect\citename{Lockhart et al., 2010}]{PICOpaper}
Lockhart M., Person M., \& Elliot J., 2010, PASP, 122, 1207

\bibitem[\protect\citename{Mousis et al., 2014}]{ProAmPaper}
Mousis O., Hueso R., Beaulieu J.P. et al., 2014, ExA 38, 1-2, pp 91-191

\bibitem[\protect\citename{Pavlov, 2012a}]{TangraWebSite}
Pavlov, H. 2012a. Tangra v3 software. Download from http://www.hristopavlov.net/Tangra3/

\bibitem[\protect\citename{Pavlov, 2012b}]{ADVGitHub}
Pavlov, H. 2012b. ADV file format development. Download from https://github.com/AstroDigitalVideo/

\bibitem[\protect\citename{Point Grey Research, 2012}]{GExSpecs}
Point Grey Research, 2012. Grasshopper Express Technical Manual. Download from http://www.ptgrey.com/support/downloads/10126/

\bibitem[\protect\citename{Roques et al., 2009}]{SolarSystemOccPaper}
Roques F., Boissel Y., Doressoundiram A. et al., 2009, EM\&P 105:201

\bibitem[\protect\citename{Sicardy et al., 2011}]{CharonOcc2008Paper}
Sicardy B, Broughton J., Bolt G. et al., 2011, AJ, 131:67

\bibitem[\protect\citename{Souza et al., 2006}]{POETSpaper}
Souza S.P., Babcock B.A., Pasachoff J.M., et al., 2006, PASP, 118, 1550

\bibitem[\protect\citename{Regester et al., 2011}]{PHOTpaper}
Regester J., Young E.F., Young L.A., Olkin C.B. et al., 2011, PASP, 123, 735

\bibitem[\protect\citename{Warner B., 1972}]{StellAngDiamPaper}
Warner B., 1972, MNRAS 158:1-3

\end{thebibliography}
\end{document}